\RequirePackage{ifpdf}
\ifpdf % We are running pdfTeX in pdf mode
\documentclass[pdftex]{sigma}
\else
\documentclass{sigma}
\fi

\begin{document}
\allowdisplaybreaks

\renewcommand{\PaperNumber}{033}

\FirstPageHeading

\ShortArticleName{Zonal Functions and Fourier Transforms}

\ArticleName{A New Form of the Spherical Expansion \\
of Zonal Functions and Fourier Transforms \\ of $\boldsymbol{{\bf
SO}(d)}$-Finite Functions}

\Author{Agata BEZUBIK~$^\dag$ and Aleksander STRASBURGER~$^\ddag$}
\AuthorNameForHeading{A. Bezubik and A. Strasburger}

\Address{$^\dag$~Institute of Mathematics, University of
Bia\l{}ystok, Akademicka 2, 15-267 Bia\l{}ystok, Poland}
\EmailD{\href{mailto:agatab@math.uwb.edu.pl}{agatab@math.uwb.edu.pl}}

\Address{$^\ddag$~Department of Econometrics and Informatics, Warsaw Agricultural University,\\
$\phantom{^\ddag}$~Nowoursynowska 166, 02-787 Warszawa, Poland}
 \EmailD{\href{mailto:strasburger@alpha.sggw.waw.pl}{strasburger@alpha.sggw.waw.pl}}

\ArticleDates{Received November 30, 2005, in f\/inal form February
17, 2006; Published online March 03, 2006}

\Abstract{This paper presents recent results obtained by the
authors (partly in collaboration with A.~D\c{a}browska) concerning
 expansions of zonal functions on Euclidean spheres into
 spherical harmonics and some applications of such expansions
 for problems involving Fourier transforms of functions with rotational symmetry.
 The method used to derive the expansion formula is based entirely
 on dif\/ferential methods and completely avoids the use of various
 integral identities commonly used in this context.
 Some new identities for the Fourier transform are
 derived and as a byproduct seemingly new recurrence relations
  for the classical Bessel functions are obtained.}

\Keywords{spherical harmonics; zonal harmonic polynomials;
Fourier--Laplace expansions; special orthogonal group; Bessel
functions; Fourier transform; Bochner identity}

\Classification{33C55; 42B10; 33C80; 44A15; 44A20}

\begin{flushright}
\begin{minipage}{7.7cm}
{\it\small This paper is dedicated to Professor Jacques Faraut on
the occasion of his 65-th anniversary.}
\end{minipage}
\end{flushright}

\section{Introduction and preliminaries on spherical harmonics}

Many parts of analysis on Euclidean spaces and in particular the
theory of spherical harmonics provide an elegant and instructive
application of group theoretical concepts to various questions of
classical function theory. It suf\/f\/ices to mention points like:
coincidence of eigenspaces of spherical Laplacian with irreducible
components with respect to the natural action of the group ${\bf
SO}(d)$ within the $L^2$ space on the sphere; the role of
classical orthogonal polynomials, i.e.\ Gegenbauer polynomials, as
reproducing kernels for the spaces of spherical harmonics of a
given degree, or more generally, as providing an explicit
construction of symmetry adapted basis functions for those spaces
(cf.~\cite{Vi}), or the connection of the Fourier transform on the
Euclidean space to the Hankel transform obtained via restriction
to ${\bf SO}(d)$-f\/inite functions and various integral
identities of the Hecke--Bochner type resulting there from.

The goal of the present paper is to present under one cover recent
developments within this circle of ideas, obtained in
\cite{me3,BDS1,BDS2} by the present authors, partly in
collaboration with A.~D\c{a}browska. The central point of this
paper is a novel form of spherical expansion of zonal functions on
Euclidean spheres which we derive by purely dif\/ferential
methods. We show how it implies certain  general formulae for the
Fourier transform of ${\bf SO}(d)$-f\/inite functions, including
recent generalizations of the famous Bochner formula and also
derive certain function theoretic consequences.

\subsection{Preliminaries and notations}
We denote by $(x\,|\, y)$ the standard (Euclidean) inner product
of points $x,y\in {\mathbb R}^d$  and by $r=|x|=(x\,|\,x)^{1/2}$ the
corresponding length (or radius) function. The unit sphere in
$\mathbb R^d$ is denoted by $S^{d-1}$ and for $x\neq0$ we shall
often write $x=r\xi$ with $\xi\in S^{d-1}$. We shall assume $d\ge
3$ and frequently use a constant $\alpha$ def\/ined by
$\alpha=(d-2)/2$. The sphere is regarded as a homogeneous space of
the group $K={\bf SO}(d)$ of (proper) rotations in ${\mathbb R}^d$
and for a point $\eta\in S^{d-1}$ its isotropy group
$K_\eta\subset K$ can be identif\/ied with the group ${\bf
SO}(d-1)$.

Let $\Delta = \sum\limits_{j=1}^{d}{ \partial^2}/{\partial x_j^2}$
denote the Laplacian in  ${\mathbb R}^d$. We let ${\mathcal P}^l
={\mathcal P}^l ({\mathbb R}^d)$ denote the space consisting of
homogeneous polynomials of degree $l$ on ${\mathbb R}^d$ and
denote the kernel of $\Delta$ in ${\mathcal P}^l$ by~${\mathcal
H}^l $. This latter space consists of harmonic and  homogeneous
polynomials of degree $l$ and its dimension is given by
\[
\dim {\mathcal
H}^l=\frac{2(l+\alpha)\Gamma(2\alpha+l)}{\Gamma(l+1)\Gamma(2\alpha+1)},
\]
where $\Gamma(z)$ denotes the Euler gamma function. Both these
spaces are invariant under the natural action of the group $K={\bf
SO}(d)$ on functions on ${\mathbb R}^d$ given by $k\cdot
f(x)=f(k^{-1}x)$ for $k\in K$, $x\in \mathbb R^d$. It is known
that this group acts irreducibly in ${\mathcal H}^l$ for each $l$,
and the representations so obtained are inequivalent for $l\neq
l'$.

The \emph{surface spherical harmonics of order} $l$ are by
def\/inition the restrictions of  elements from~${\mathcal H}^l$
to the unit sphere $S^{d-1}$, and since the restriction map
commutes with the action of rotations the spaces of surface
spherical harmonics of any f\/ixed order are invariant and
irreducible under $K$.

We endow ${\mathcal P}^l={\mathcal P}^l({\mathbb R}^d)$ with an
inner product def\/ined by the formula
\begin{gather} \label{inner-prod}
\langle P\,|\, Q\rangle := \int_{S^{d-1}}
P(\xi)\overline{Q(\xi)}\,d\sigma(\xi),
\end{gather}
where $d\sigma(\cdot)$ is the Euclidean surface measure on the
unit sphere $S^{d-1}$ normalized so that the total area of the
sphere is 1. In fact, the integral on the right hand side extends
to the natural inner product on $L^2(S^{d-1}, d\sigma)$ and it is
known that the spaces of spherical surface harmonics of
dif\/ferent orders are orthogonal to each other with respect to
this inner product.

We recall that a function $f$, def\/ined on the unit sphere
$S^{d-1}$, is said to be \emph{a zonal function (relative to a
point $\eta\in S^{d-1}$)} if it is invariant with respect to the
isotropy group $K_\eta$ of $\eta$. Any such function depends in
fact on the inner product $(\xi \,|\, \eta)$ only, and so there
exists a function $\phi$ def\/ined on the closed unit interval
$[-1,1]\subset {\mathbb R}$ so that
\begin{gather}\label{eqn:profile}
f(\xi)= \phi((\xi\,|\, \eta)),\qquad  \text{for all}\ \xi\in
S^{d-1}.
\end{gather}
We shall call $\phi$ the prof\/ile function of $f$.

\subsection{The canonical decomposition of homogeneous polynomials}

It is well known that every homogeneous polynomial of degree $l$
can be represented as a sum of products of harmonic homogeneous
polynomials with even powers of the radius,
cf.~equation~(\ref{can-decomp2}) below, and this decomposition is
usually called the canonical decomposition of homogeneous
polynomials. While the general form of the canonical
decomposition, i.e.\ equation~(\ref{can-decomp2}), 
is stated in all sources concerned with this matter, the explicit formula for harmonic
components of a~given homogeneous polynomial is seldom quoted. The
only source known to us, where the formula can be found, is the
paper \cite{Lu} of Lucquiaud. We record it here since it is
essential for considerations to follow, in particular, in
establishing formula (\ref{expansions3}) giving expansion of
elementary zonal polynomials of the form $(x\,|\,\eta)^l$, where
$\eta \in S^{d-1}$ is an arbitrary unit vector.

\begin{theorem}[The canonical decomposition] \label{can-decomp}
The space ${\mathcal P}^l$ decomposes orthogonally as the sum
${\mathcal P}^l = \oplus_{k=0}^{[l/2]} r^{2k} {\mathcal H}^{l-2k}$
and the decomposition is invariant with respect to the action of
the group ${\bf SO}(d)$. Explicitly, every polynomial $P\in
{\mathcal P}^l$ may be written as
\begin{gather}\label{can-decomp2}
P=\sum_{k=0}^{[l/2]}r^{2k}h_{l-2k}(P), \qquad \text{where}\quad
h_{l-2k}(P)\in {\mathcal H}^{l-2k}
\end{gather}
are called the harmonic components of $P$ and are given by
\begin{gather}\label{eqn:harmon_proj}
h_{l-2k}(P)= \sum^{[l/2]-k}_{j=0}  e^{l}_{\,j}(k) r^{2j}
\Delta^{k+j} P
\end{gather}
with the coefficients $e^{l}_{\,j}(k)$ determined by
\begin{gather}\label{eqn:harm_coeff}
e^{l}_{\,j}(k) = (-1)^j\frac
{(\alpha+l-2k)\Gamma(\alpha+l-2k-j)}{4^{k+j}k!j!\Gamma(\alpha+l+1-k)}.
\end{gather}
The maps $P\mapsto r^{2k} h_{l-2k}(P)\in P^l$ are
projections  onto ${\bf SO}(d)$-irreducible subspaces of
${\mathcal P}^l$ commuting with the action of the group ${\bf
SO}(d)$.
\end{theorem}

\section{Expansions of zonal functions}

\subsection{Elementary zonal functions and reproducing kernels}

To proceed, we need to recall an explicit formula for the
Gegenbauer polynomial $C^\alpha_l$ of degree $l$ and index
$\alpha$, given for example in \cite[Section 5.3.2]{MOS} or
\cite[Chapter 9.2]{Waw}, which reads as follows:
\begin{gather*}
C_l^\alpha(t) = \sum_{j=0}^{[l/2]}(-1)^j
\frac{\Gamma(\alpha+l-j)}{\Gamma(\alpha)\Gamma(j+1)\Gamma(l+1-2j)}
(2t)^{l-2j}.
\end{gather*}

Applying the formulae for the canonical  decomposition given in
(\ref{eqn:harmon_proj}) and (\ref{eqn:harm_coeff}) to elementary
zonal polynomials $(x\,|\, \eta)^l$, where  $\eta\in S^{d-1}$ and
$l$ is a nonnegative integer, we obtain
\begin{gather}\label{expansions3}
(x\,|\, \eta)^l = 2^{-l}\Gamma(\alpha)\Gamma(l+1)|x|^l
\sum_{k=0}^{[l/2]} \frac{(\alpha+l-2k)}{k!\Gamma(\alpha+l+1-k)}
C_{l-2k}^\alpha\bigl((\xi\,|\, \eta)\bigr), \qquad x=|x|\xi.
\end{gather}

In particular, the spherical harmonic obtained  by restricting to
the unit sphere the harmonic component $h_l(P_{\eta})$ of the
highest degree of $P_\eta(x)=(x\,|\,\eta)^l$, which is given by
the formula
\begin{gather*}%\label{eqn:first_harm_proj}
h_l(P_\eta)(\xi)=2^{-l}\Gamma(\alpha)\Gamma(l+1)\frac{(\alpha
+l)}{\Gamma(\alpha +l+1)}C_l^{\alpha}((\xi\,|\,\eta)),
\end{gather*}
plays an important role in the group representation theoretic
interpretation of the decomposition~(\ref{can-decomp}). With
normalization given by
\begin{gather*} %\label{zonal}
Z^l_\eta(\xi)=
\bigl[C_l^{\alpha}(1)\bigr]^{-1}C_l^{\alpha}((\xi\,|\, \eta))
\end{gather*}
it satisf\/ies
\[
\dim{\mathcal H}^l\int_{S^{d-1}} Z^l_\eta(\xi)P(\xi)\,d\sigma(\xi)
= P(\eta).
\]
Because of this property, $Z^l_\eta(\xi)$ is called the
reproducing kernel for the space ${\mathcal H}^l$.  Moreover, it
is uniquely (up to a scalar multiple) determined by the property
of being invariant under the action of the isotropy subgroup
$K_\eta$ of the point $\eta\in S^{d-1}$.

\subsection[A differential formula for expansions of smooth zonal functions]{A dif\/ferential formula for expansions of smooth zonal functions}

Now recall \cite{Mue, Vi} that every square integrable function on
the sphere can be written as a series of spherical harmonics (the
Fourier--Laplace expansion). For zonal functions this expansion,
thanks to an easy group representation theoretic argument, reduces
to
\begin{gather*}%\label{eqn:expansion1}
  f(\xi)= \sum_{m=0}^{\infty} f_m Z^m_\eta(\xi), \qquad  \text{where}\quad
  f_m = \dim{\mathcal H}^m\int_{S^{d-1}}f(\xi)Z^m_\eta(\xi)\,d\sigma(\xi).
\end{gather*}
By taking into account the invariance under $K_\eta$ of the
integrand and equation (\ref{eqn:profile}),  the coef\/f\/icients
may be expressed as integrals
\begin{gather*}%\label{eqn:integrals}
f_m =
\frac{(\alpha+m)\Gamma(\alpha)}{\sqrt{\pi}\Gamma(\alpha+1/2)}
 \int_{-1}^{1}\phi(t)C_m^\alpha(t)\big(1-t^2\big)^{\alpha-1/2}\,dt,
\end{gather*}
what reduces the problem of spherical expansion of zonal functions
to the expansion of prof\/ile functions $\phi$ with respect to the
(orthogonal) system of Gegenbauer polynomials.  The following
result shows that the coef\/f\/icients $f_m$ of the expansion can
also be expressed in terms of the coef\/f\/icients of the Taylor
expansion of the prof\/ile function $\phi$, provided the latter
satisf\/ies suitable regularity assumptions.

\begin{theorem}\label{expan_of_f}
Assume $\phi:[-1,1] \rightarrow {\mathbb C}$ has the Taylor
expansion $\phi(t)= \sum\limits_{m=0}^{\infty}
\frac{\phi^{(m)}(0)}{m!}t^m$ that is absolutely convergent on the
closed interval $[-1,1]$, and let $f(\xi)=\phi((\xi \,|\, \eta))$
be the zonal function on the sphere $S^{d-1}$ corresponding to
$\phi$. Then the spherical Fourier expansion of $f(\xi)$ is given
by
\begin{gather}\label{exp_f}
f(\xi)=\Gamma(\alpha+1) \sum_{m=0}^{\infty} f_m \dim{\mathcal H}^m
Z^{m}_{\eta}(\xi),
\end{gather}
where the coefficients of the expansion can be expressed as
\begin{gather*}%\label{exp_coeff}
f_m = \sum_{k=0}^{\infty}
\frac{\phi^{(m+2k)}(0)}{2^{m+2k}k!\Gamma(\alpha+m+k+1)}.
\end{gather*}
\end{theorem}
A detailed proof of this result is contained in the forthcoming
paper \cite{BDS2} of the authors, and here we shall present its
main line only. It consists in substituting the expansion formula
(\ref{expansions3}) into the Taylor series of $\phi$ and
rearanging terms so that to group together the terms corresponding
to the Gegenbauer polynomials of a given degree. This procedure
requires some estimates on the coef\/f\/icients which assure the
absolute convergence of the double series, which are presented in
detail in~\cite{BDS2}.

Below we brief\/ly present two applications of this expansion for
obtaining new derivations of some classical results.

\subsection{The plane wave expansion}
A well known and very useful instance of the expansion
\eqref{exp_f} is the so called plane wave expansion, giving a
representation of the exponential function $e^{i(x|\eta)}$ as a
series of zonal harmonic polynomials. The expansion involves the
Bessel functions of the f\/irst kind of order $\nu \in {\mathbb
C}$ with $\mathop{\rm Re}\nolimits{\nu}>-1$, which are given as
\begin{gather}\label{Bessel}
J_\nu(t)= \left(\frac{t}{2}\right)^\nu\sum_{k=0}^{\infty}
\frac{(-1)^k}{\Gamma(k+1)\Gamma(k+\nu+1)}\left(\frac{t}{2}\right)^{2k},
\qquad t\in {\mathbb C}.
\end{gather}
For some purposes the results are better expressed with the aid of
the so called spherical Bessel functions, which are given as
\begin{gather*}%\label{bessel}
j_\nu(t)= \Gamma(\nu+1)\left(\frac{t}{2}\right)^{-\nu} J_\nu(t).
\end{gather*}
We point out that all classical proofs of this formula known to us
are obtained by applying certain integral identities of the
Hecke--Bochner type, as in \cite{far,Mue}. With the use of
\eqref{exp_f} we get it by direct dif\/ferentiation of the
exponential $e^{irt}$ and comparison with the power series
expansion of the Bessel function \eqref{Bessel}.

\begin{corollary}\label{d>2}
For an arbitrary unit vector $\eta\in S^{d-1}\subset {\mathbb
R}^d$ and $x\in {\mathbb R}^d$, the plane wave $e^{i(x|\eta)}$
admits the following expansion
\begin{gather*}%\label{pl_w_exp_3}
e^{i(x|\eta)}
 = \sum_{m=0}^{\infty}i^m \dim {\mathcal H}^m \frac{\Gamma(\alpha+1)}{\Gamma(\alpha+m+1)}
\left(\frac{r}{2}\right)^m j_{\alpha+m}(r)Z^m_\eta(\xi), \qquad
x=r\xi, \quad \xi\in S^{d-1}.
\end{gather*}
The series converges absolutely on each sphere of radius $r$ and
uniformly with respect to $\xi,\eta\in S^{d-1}$.
\end{corollary}

\subsection{The generating function of Gegenbauer polynomials}

Take a f\/ixed element $x\neq 0$ from the unit ball in ${\mathbb
R}^d$ and write $x=r\eta$, $\eta\in S^{d-1}$. The function
\[
S^{d-1}\ni \xi \mapsto |\xi- x|^{-2\alpha}=
\left(1-2r(\xi\,|\,\eta) +r^2\right)^{-\alpha}
\]
is clearly a zonal function with pole at $\eta$ corresponding to
the prof\/ile function $(1-2rt +r^2)^{-\alpha}$. The expansion
resulting from \eqref{exp_f} has the form
\begin{gather*}%\label{eqn:generating}
  (1-2r(\xi\,|\,\eta) +r^2)^{-\alpha}= \sum_{m=0}^{\infty} \frac{\Gamma(\alpha+m)}{\Gamma(2\alpha)\Gamma(m+1)} r^m Z^m_\xi(\eta)
\end{gather*}
and can be reduced to the familiar formula for the generating
function of Gegenbauer polynomials
\begin{gather*}
\left(1-2rt +r^2\right)^{-\alpha}= \sum_{m=0}^{\infty}r^m
C^{\alpha}_m(t),
\end{gather*}
as given e.g.\ in \cite{AA,far,MOS}.

\section[The Fourier transforms of ${\bf SO}(d)$-finite functions]{The Fourier transforms of
$\boldsymbol{{\bf SO}(d)}$-f\/inite functions}

In this section we shall present some applications of the plane
wave expansion given in Corollary~\ref{d>2} to problems of Fourier
analysis in Euclidean space. The subject is related to the well
known Bochner formula (c.f.\ e.g.\ \cite{AA, Mue}), which
describes the restriction of the Fourier transform to  the space
of functions of the form  $f(r)P(x)$, where $f$ is a radial
function and  $P\in {\mathcal H}^l$ a   homogeneous harmonic
polynomial. The result is again a product of the same harmonic
polynomial $P$ with a radial function, which is expressed as the so
called Hankel transform of the original radial factor and is given
by the formula~(\ref{Hankel_trans}) below. The result we present
in Corollary~\ref{Bochner_id} generalizes that relation for the
case of arbitrary homogeneous polynomials and is a combination of
results obtained by the authors in \cite{me3,BDS1} and by F.J.~Gonzalez-Vieli in \cite{Gon}.

\subsection[The case of ${\bf SO}(d)$-finite functions on the sphere]{The case
of $\boldsymbol{{\bf SO}(d)}$-f\/inite functions on the sphere}

We take the Fourier transform of suitable regular (e.g.
$L^1({\mathbb R}^d)$) functions on ${\mathbb R}^d$ as def\/ined by
\[
{\mathcal F} f(x)=(2\pi)^{-\frac{d}{2}}\int_{{\mathbb
R}^d}e^{i(x\,|\, y)}f(y)\,dy,\qquad x\in {\mathbb R}^d.
\]
Observe that the above def\/inition of the Fourier transform also
makes sense in the case when $f$ is a function def\/ined on the
unit sphere --- in this case it can be regarded as the Fourier
transform of a measure supported on the sphere. Especially
interesting is the case, when the measure on the sphere comes from
restricting a homogeneous polynomial to the sphere, since this way
we obtain a so-called ${\bf SO}(d)$-f\/inite measure (since its
${\bf SO}(d)$-translates span a f\/inite dimensional subspace).
\begin{theorem}
\label{Theo:FJG} If $P\in {\mathcal P}^l$, then the Fourier
transform of the measure $P(\xi) d\sigma(\xi)$ with support on the
unit sphere $S^{d-1}$ is given by the following equivalent
formulae
\begin{gather}
{\mathcal F}(P)=\int_{S^{d-1}}e^{i(x\,|\,\eta)}P(\eta)\,
d\sigma(\eta)\nonumber \\
\phantom{{\mathcal F}(P)}{} =
\biggl(\frac{i}{2}\biggr)^l\sum_{k=0}^{[l/2]}
\frac{(-1)^k2^{2k}\Gamma(\alpha+1)}{\Gamma(\alpha+l+1-2k)}
j_{\alpha+l-2k}(|x|)h_{l-2k}(P)(x)
\label{eqn:FT3} \\
\phantom{{\mathcal F}(P)}{}=\left(\frac{i}{2}\right)^l
\sum_{k=0}^{[l/2]}
\frac{(-1)^{k}\Gamma(\alpha+1)}{k!\Gamma(\alpha+l+1-k)}
j_{\alpha+l-k}(|x|)(\Delta^k P)(x) \label{eqn:FJG}
\end{gather}
with $h_{l-2k}(P)$ denoting the harmonic components of $P$ as in
equation~\eqref{can-decomp2}.
\end{theorem}

The formula \eqref{eqn:FJG} has been derived by 
F.J.~Gonzalez~\cite{Gon} and the equivalence of those two forms was
observed by the present authors and A. D\c{a}browska in
\cite{BDS1}.

Theorem \ref{Theo:FJG} implies the following interesting function
theoretical corollary. By comparing the two expressions
(\ref{eqn:FT3}) and (\ref{eqn:FJG}) from this theorem one may
derive the following multi-step recurrence relation for spherical
Bessel functions.

\begin{corollary}\label{Coro:multistep}
If $\alpha\ge0$ is a half odd integer (or an integer) then for any
$l\in {\mathbb Z}_+$ the following relations hold
\begin{gather*}%\label{eqn:mult}
 j_{\alpha+l-s}(r)=
\sum_{k=0}^{s} \frac{s!\,
\Gamma(\alpha+l+1-s)\Gamma(\alpha+l-k-s)}{k!(s-k)!\Gamma(\alpha+l+1-k)\Gamma(\alpha+l-2k)}
\left(\frac{r}{2}\right)^{2(s-k)}j_{\alpha+l-2k}(r)
\end{gather*}
for $s=1,\ldots, [l/2]$.

In terms of the Bessel functions of the first kind, this is
expressed by
\begin{gather}\label{Bess:mult}
\frac{1}{s!}\left(\frac{2}{r}\right)^{s}J_{\alpha+l-s}(r)=
\sum_{k=0}^{s}
\frac{\Gamma(\alpha+l-k-s)\Gamma(\alpha+l+1-2k)}{k!(s-k)!\Gamma(\alpha+l+1-k)\Gamma(\alpha+l-2k)}J_{\alpha+l-2k}(r).
\end{gather}
\end{corollary}

It is interesting to note that these latter relations
 unify several classical relations satisf\/ied by Bessel functions
 like  \cite[equation~(4.6.11)]{AA} or \cite[Chapter~3.2.2]{MOS}.
In fact, taking into account the relation
\[
J_{-1/2}(t)=\left(\frac{1}{2}\pi t\right)^{-1/2}\cos{t},
\]
which follows directly from the expansion (\ref{Bessel}), and  the
familiar recurrence relations satisf\/ied by~$J_\nu(t)$, namely
\begin{gather*}%\label{recurrence}
 \left(\frac{1}{t}\frac{d}{d t}\right)^l (t^{-\nu}J_\nu(t))=
(-1)^l t^{-\nu-l}J_{\nu+l}(t)
\end{gather*}
one can derive from \eqref{Bess:mult} the following relations.

\begin{corollary}[Finite expansions of Bessel functions]
The Bessel functions of integer order satisfy the relation
\begin{gather*}
\frac{1}{n!}\left(\frac{2}{t}\right)^{n}J_n(t)=\sum_{k=0}^{n}\epsilon_k\frac{1}{(n+k)!(n-k)!}J_{2k}(t),
 \quad \textrm{where} \quad
\epsilon_k=\left\{\begin{array}{ll}
1, & k=0\\
2, & k=1,2,3\ldots,
\end{array}\right.
\end{gather*}
while those of half odd integer order satisfy
\begin{gather*}
J_{n+1/2}(t)= \sqrt{\frac{2}{\pi x}} \left\{
\sin(t-n\pi/2)\sum_{k=0}^{[n/2]}
\frac{(-1)^k(n+2k)!}{(2k)!(n-2k)!(2t)^{2k}}\right.  \\
 \left. \phantom{J_{n+1/2}(t)=}{}+
\cos(t-n\pi/2)\sum_{k=0}^{[(n-1)/2]}\frac{(-1)^k(n+2k+1)!}{(2k+1)!(n-2k-1)!(2t)^{2k+1}}
\right\}.
\end{gather*}
\end{corollary}

\subsection{The generalized Bochner formula and the Hankel transform}
For our last topic recall that for suitable functions on the
nonnegative real axis ${\mathbb R}_+$, say, with $\phi$ belonging
to the Schwartz space ${\mathcal S}({\mathbb R})$, one def\/ines
the Hankel transform by the formula
\begin{gather}\label{Hankel_trans}
H_\nu(\phi)(t)  =  \frac{2^{-\nu}}{\Gamma(\nu+1)}\int_0^\infty
\phi(s)j_\nu(st)s^{2\nu+1}ds.
\end{gather}
Theorem \ref{Theo:FJG} above now immediately implies that the
generalized Bochner formula which was previously obtained in
\cite{me3} can also be expressed by a pair of equivalent formulae.

\begin{corollary}[Generalized Bochner identity]\label{Bochner_id}
If $P\in {\mathcal P}^l$, then the Fourier transform of the
function $f(|x|)P(x)$ is given by the following expressions:
\begin{gather*}
(2\pi)^{-\frac{d}{2}}\int_{{\mathbb R}^{d}}f(|x|)P(x)e^{i(y| x)}dx
 =  i^l\sum_{k=0}^{[l/2]}
(-1)^k H_{\alpha+l-2k}(s^{2k}f)(|y|)h_{l-2k}(P)(y) \\
\phantom{(2\pi)^{-\frac{d}{2}}\int_{S^{d-1}}f(|x|)P(x)e^{i(y|
x)}dx}{} =  i^l\sum_{k=0}^{[l/2]} \frac{(-1)^k}{2^k k!}
H_{\alpha+l-k}(f)(|y|)\Delta^{k}P(y)
\end{gather*}
\end{corollary}

This implies in turn the following:
\begin{corollary}[Periodicity relation for the Hankel transform]
For any $\phi\in{\mathcal S}({\mathbb R})$ with~$\alpha$ and $l$
satisfying
 conditions of Corollary {\rm \ref{Coro:multistep}} above, the Hankel transform satisfies the following relation
\begin{gather*}
H_{\alpha+l}(\phi)(t) = 2(\alpha+l-1)H_{\alpha+l-1}(\phi)(t) -
H_{\alpha+l-2}\left(s^{2}\phi\right)(t).
\end{gather*}
\end{corollary}

\section{Conclusion}
In this paper we have demonstrated that dif\/ferential identities
for homogeneous polynomials like those implied by equations
\eqref{can-decomp2}--\eqref{eqn:harm_coeff} can be ef\/fectively
used for solving problems in harmonic analysis, which so far have
been approached by means of integral identities of the type of
Hecke--Bochner formula. In our opinion other possibilities of
using that approach
 should certainly be further explored.

\subsection*{Acknowledgements}
The results contained in this paper were presented at the
conference Symmetry in Nonlinear Mathematical Physics in Kyiv,
June 20--26, 2005 and also at the Seminar Sophus Lie
 in Nancy, June 10, 2005. We thank the organizers of those
 meetings for enabling us to present these results there.
We are also obliged to the referees for remarks which, as we hope,
enabled us to improve the presentation in the paper. In
particular, the reference~\cite{CF} was indicated by the referee.

\LastPageEnding

\end{document}